\documentclass[a4paper,11pt]{article}
\pdfoutput=1 % if your are submitting a pdflatex (i.e. if you have
\usepackage{jinstpub} % for details on the use of the package, please
\title{Status and perspectives of the PETALO project}

\author[a]{Paola Ferrario}

\affiliation[a]{Donostia International Physics Center (DIPC), Paseo Manuel Lardizabal 4, Donostia-San Sebasti\'an, E-20018, Spain and Basque Foundation for Science (IKERBASQUE), Bilbao, E-48013, Spain.}

\emailAdd{paola.ferrario@dipc.org}

\abstract{PETALO (Positron Emission Tof Apparatus with Liquid xenOn) is a novel concept for positron emission tomography scanners, which uses liquid xenon as a scintillation medium and silicon photomultipliers as a readout. The large scintillation yield and the fast scintillation time of liquid xenon, as well as its scalability, makes it an excellent candidate for PET scanners with Time-of-Flight measurements, especially for total-body machines. A first prototype of PETALO, devoted to demonstrate the potential of the concept, measuring the energy and time resolution and to test technical solutions for a complete ring is fully operational. The prototype consists of an aluminum box filled with liquid xenon, with two arrays of SiPMs on opposite sides facing the xenon. A $\beta$+ emitter source generating 511-keV pairs of gammas is placed in a central port and the SiPMs record the scintillation light produced by the gamma interactions, allowing for the reconstruction of the position, the energy and the time of the interactions.}


\collaboration[c]{on behalf of PETALO collaboration}

\proceeding{LIDINE 2021: LIght Detection in Noble Elements\\
  14-18 September 2021\\
  University of California San Diego}

\begin{document}
\maketitle
\flushbottom

\section{Introduction}
\label{sec:intro}

Positron Emission Tomography (PET) is an imaging technique which scans
the metabolic activity of the body and is widely used to monitor
tumours, neurodegenerative disease like Alzheimer's, epilepsy or
cardiovascular diseases. Glucose molecules, modified with a $\beta$+
emitting atom like $^{18}$F, are injected into the body of the
patient as they tend to concentrate in regions of higher metabolic
activity. The positron emitted in the decay annihilates with one of
the electrons of the surrounding atoms, creating two high energy
photons of 511 keV emitted at, or close to, $180^\circ$ to each
other. These photons can be detected in time coincidence in a ring of
scintillators placed around the patient. Each coincidence forms a line and
the intersection of several of these lines can be used to reconstruct
the image of the region where the radioisotope concentrates. If, in
addition to position reconstruction, one is able to measure the interaction times of the photons with high
precision (a Time-Of-Flight measurement), a segment on the line can be
identified as having higher probability of containing the origin of
the gammas. This capability results in an improvement in the quality
of the image proportional to the time resolution achieved in the
scanner~\cite{a}. As a consequence, smaller lesions can be imaged for
the same amount of received dose of radioisotope or, conversely, an
image of the same quality can be obtained in a shorter time, thus
reducing the dose received by the patient.

\section{A liquid xenon PET scanner}
\label{sec:fullbody}

Xenon is a noble gas, which produces approximately 60 scintillation
photons per keV of deposited energy when ionizing radiation interacts
with its atoms~\cite{b}. When a 511 keV gamma interacts in liquid
xenon (LXe) (through photoelectric interaction or Compton scatter) it produces
secondary electrons which in turn propagate for a short distance in
the liquid, ionizing and exciting the medium. The scintillation signal
is primarily produced by the de-excitation of the xenon atoms to
dimers which in turn decay producing ultraviolet (VUV) radiation
peaked at 178~nm with characteristic decay constants of 2.2~ns
(singlet mode) and 27~ns (triplet mode) \cite{kubota}. There is also a slower
contribution (tens of nanoseconds) from the electron-ion
recombination. In its liquid phase xenon has a reasonably high density
(2.98~g/cm$^3$)~\cite{c} and as a result an attenuation length of 3.7~cm for 511-keV gammas~\cite{d}  and a long Rayleigh scattering length compared to typical size of a scanner (36.4 cm)~\cite{e}, which makes it suitable for PET applications. 

The PETALO concept~\cite{f,g} proposes a PET scanner based on the use of
liquid xenon, with silicon photomultipliers (SiPMs) to read out the
xenon scintillation light only. Such a scanner can provide excellent
performance in terms of energy, spatial and time resolution. SiPMs are
widely used to read scintillation light produced in PET scanners due
to their fast response, high granularity, large gain and compatibility
with magnetic fields (allowing for combined PET-NMR scanners). In the
case of liquid xenon, a further advantage is that they have an extremely
low dark count rate at cryogenic temperatures. Moreover, since xenon
is a continuous medium, the scanner could be made of a single,
continuous volume, instead of thousands of small crystals. This
simplifies the construction and lowers costs, especially for
total-body PET scanners which have axial lengths of up to 2~m. The reconstruction of the interaction points would also benefit from reduced border effects and the purity of the liquid can be kept to the desired level through purification in gas phase, therefore ensuring the highest possible scintillation yield.

Recent Monte-Carlo studies suggest that a total-body PET based on
liquid xenon can reach an energy resolution of 15-17\% FWHM for
511-keV gammas which is of the same order as that of current scanners
based on crystals. This can be achieved in conjunction with a spatial
resolution for the interaction point of the gammas in the LXe of the order of 1 mm and a time resolution of a few hundred ps FWHM~\cite{h}.

\section{First prototype}
\label{sec:proto}

\begin{figure}[htbp]
\centering 
\includegraphics[width=.6\textwidth, trim=10cm 10cm 30cm 10cm, clip=true]{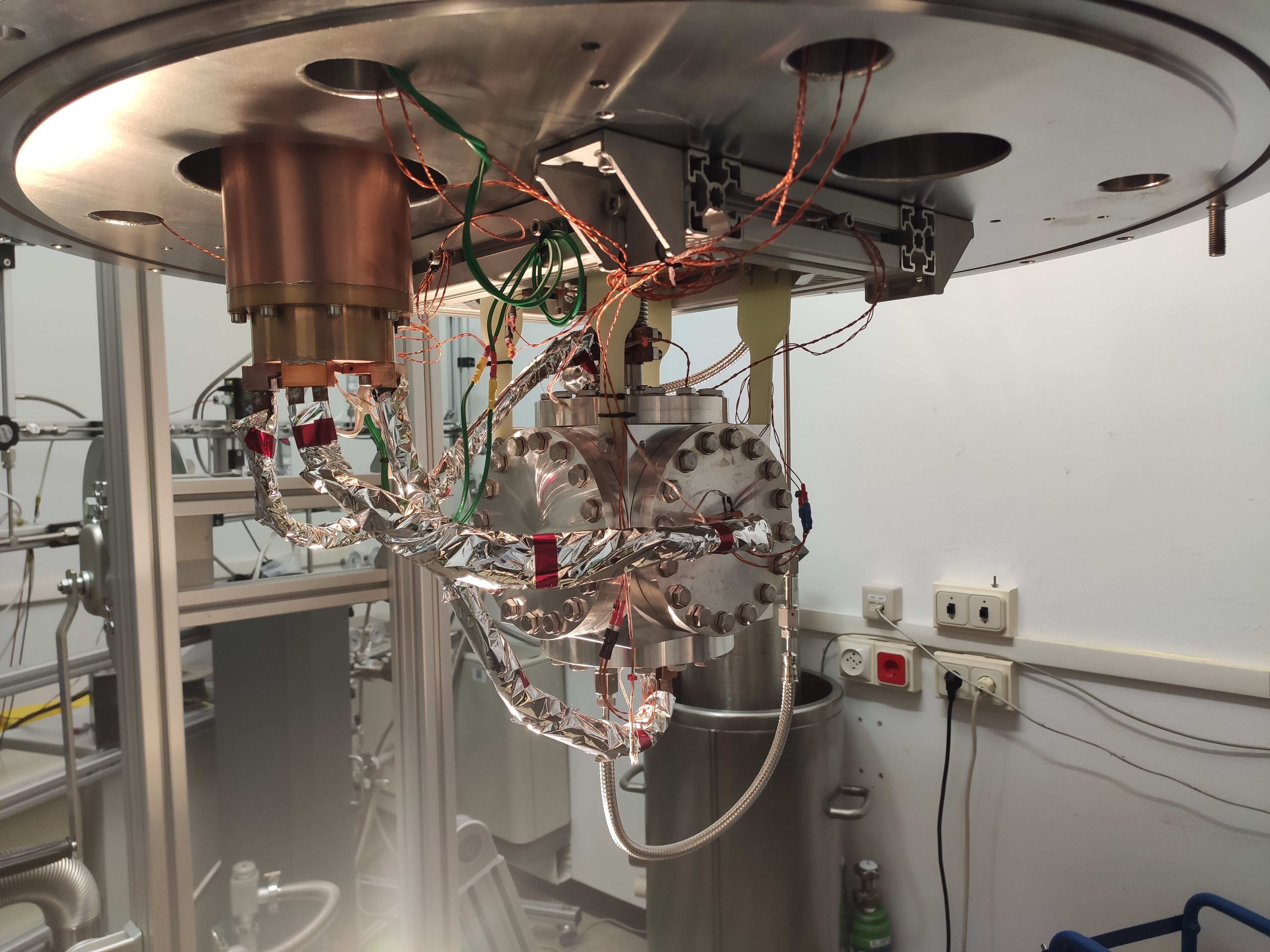}
\qquad
\includegraphics[width=.6\textwidth, trim=0cm 0cm 5cm 0cm, clip=true]{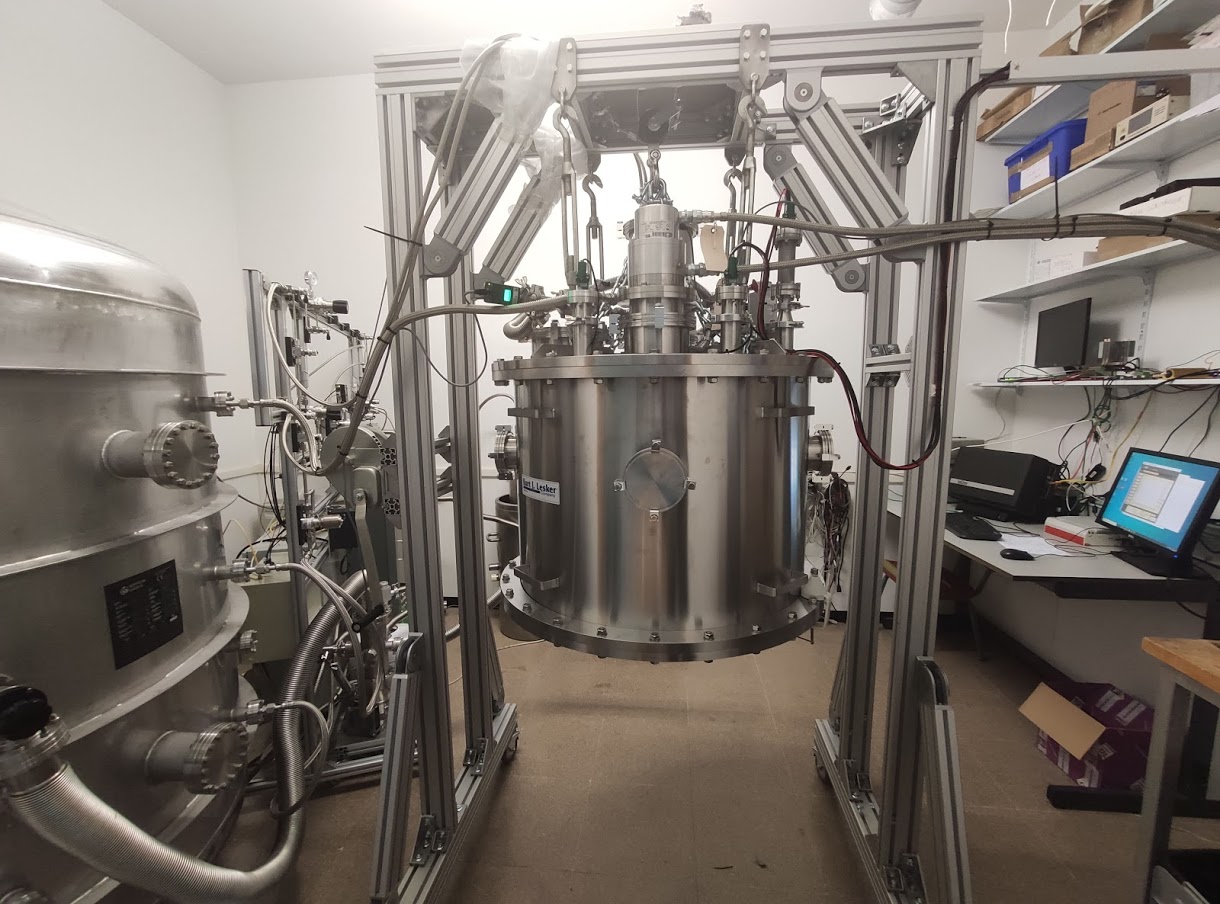}
\caption{\label{fig:box} Top: aluminum box, containing the LXe, connected to the cold head through custom-made thermal links. Bottom: view of the whole system, which comprises the vacuum vessel, the cold head, the recovery tank and the gas system.}
\end{figure}

A first prototype of PETALO has been built and is currently operating
at Instituto de F\'isica Corpuscular, in Valencia (Spain). It consists
of an aluminum cube of 10 cm side, filled with liquid xenon, kept in a
vacuum vessel for thermal insulation. The xenon is continuously
recirculated in gas phase by a double diaphragm compressor and is
purified by passing through a hot getter to remove impurities such as
nitrogen, water and oxygen, which can quench xenon scintillation
light \cite{impurities}. The cube is cooled directly by a CH-110 cold head from Sumitomo
via a set of custom-made copper thermal links. The cold head can reach
a temperature of -240 degrees Celsius using a HC-4E helium
compressor. The gas enters and exits the cube through a heat
exchanger, where the cold gas on the way out contributes to the
initial cooling of the warm gas on the way in, maximising cooling
efficiency. In the event of overpressure in the cube, the gas is
evacuated to a recovery tank, which can stand up to 3 bar of pressure,
through a release valve. In normal operation, recovery of the xenon is
achieved via the gas system lines to a cryogenic bottle. Figure~\ref{fig:box} shows the LXe container, together with the cold head and the thermal links and the fully assembled system.

\begin{figure}[htbp]
\centering 
\includegraphics[width=.6\textwidth]{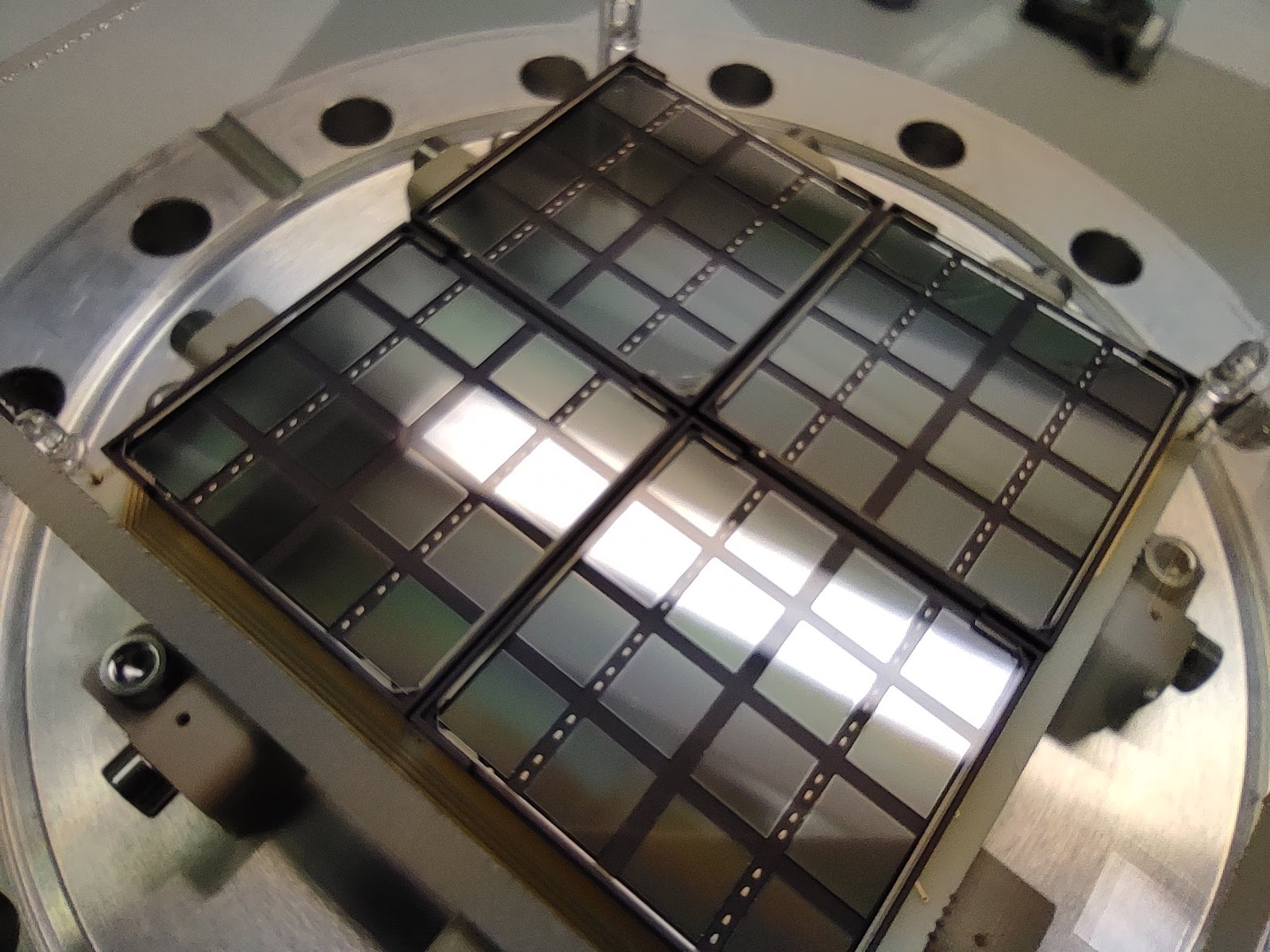}
\caption{\label{fig:sipms} Array of 64 Hamamatsu  S15779 SiPMs, sensitive to the VUV light with a photodetection efficiency of around 30\%.}
\end{figure}

The prototype is instrumented with two planes of VUV-sensitive SiPMs, mounted on opposite sides of the cube. The SiPMs have an active area of
$6\times6$~mm$^{2}$ and are arranged in arrays of $8\times8$ sensors
in each plane, facing the LXe (see figure~\ref{fig:sipms}). A
$^{22}$Na calibration source is inserted through a port in the middle
of the xenon volume. When the 511-keV gammas emitted by the positron
annihilation interact with LXe, the emitted scintillation light is
detected by the two arrays of SiPMs. The detected light is used to
determine the energy, the position of interaction and the arrival time
of the gammas. Read-out takes place via two TOFPET2 ASICs~\cite{i, l},
which integrate the charge and provide a fast timestamp. The
acquisition system has been custom designed with the aim of being
scalable to larger prototypes and, finally, to a real PET scanner. It
makes use of a novel, modular scheme with synchronization over data
links, which can be expanded to any number of channels with low
hardware cost and no extra channels or signals. Moreover, a
compression system that cuts down the amount of data sent to the
servers and inline preprocessing of the information allows for the
reduction of the speed required in the data links, thus decreasing
complexity. The DAQ is made of two boards: on the one hand, a front-end adapter 
reads the signal from the ASICs and sends it to the processor. It is also in charge of the control of the
temperature sensors, the SiPMs read-out, and the clock
distribution among ASICs. On the other hand, a data processor
distributes configuration and data between the computer and the ASICs as well as manages clock synchronization. The signals of the
sensors are extracted through a set of custom-made feedthroughs, which
also support the SiPM boards, reducing the space needed; they are
filled with high performance cryogenic epoxy, which is stable down to -200 degrees.

%\begin{figure}[htbp]
%\centering 
%\includegraphics[width=.8\textwidth, angle=180]{img/daq.jpg}
%\caption{\label{fig:daq} Acquisition system of the first PETALO prototype. The right board is the front-end adapter, directly connected to the two ASICs, while the left board is the data processor.}
%\end{figure}

\section{Conclusions}
\label{sec:concl}

The first prototype of the PETALO concept has been built and is fully
operational at the Instituto de Física Corpuscular, in Valencia
(Spain). The commissioning phase has shown system stability, with more
than two months of continuous operation without loss of xenon or any
major issues. Recently, data have started being taken with a $^{22}$Na
calibration source. Measurements of the energy and time resolution of
the system are underway and results will be published soon.

\acknowledgments
This work was supported by the European Research Council under grant ID 757829.


\end{document}